\documentclass[aps,prl,twocolumn,superscriptaddress,showpacs]{revtex4}

\usepackage{graphicx}
\usepackage{amssymb}
\usepackage{amsmath}
\usepackage{latexsym}
\usepackage{ulem}
\usepackage{color}
\usepackage{dcolumn}
\usepackage{subfigure}
\usepackage[T1]{fontenc}
\usepackage[utf8]{inputenc}
\usepackage[english,frenchb]{babel}
\usepackage[colorlinks=true,linkcolor=blue,citecolor=blue]{hyperref}%
\usepackage[squaren,Gray,cdot]{SIunits}

\begin{document}

\title{Unified modelling of the thermoelectric properties in SrTiO$_{3}$}

\author{G. Bouzerar}
\email[E-mail:]{georges.bouzerar@univ-lyon1.fr}
\affiliation{Univ Lyon, Universit\'e Claude Bernard Lyon 1, CNRS, Institut Lumi\`ere Mati\`ere, F-69622, LYON, France}
\author{S. Th\'ebaud}
\affiliation{Univ Lyon, Universit\'e Claude Bernard Lyon 1, CNRS, Institut Lumi\`ere Mati\`ere, F-69622, LYON, France}
\author{Ch. Adessi}
\affiliation{Univ Lyon, Universit\'e Claude Bernard Lyon 1, CNRS, Institut Lumi\`ere Mati\`ere, F-69622, LYON, France}
\author{R. Debord}
\affiliation{Univ Lyon, Universit\'e Claude Bernard Lyon 1, CNRS, Institut Lumi\`ere Mati\`ere, F-69622, LYON, France}
\author{M. Apreutesei}
\affiliation{INL, Site Ecole Centrale de Lyon, 36, Avenue Guy de Collongue, 69134 Ecully
France}
\author{R. Bachelet}
\affiliation{INL, Site Ecole Centrale de Lyon, 36, Avenue Guy de Collongue, 69134 Ecully
France}  
\author{S. Pailh\`es}
\affiliation{Univ Lyon, Universit\'e Claude Bernard Lyon 1, CNRS, Institut Lumi\`ere Mati\`ere, F-69622, LYON, France}

\begin{abstract}
Thermoelectric materials are opening a promising pathway to address energy conversion issues governed by a competition between thermal and electronic transport. Improving the efficiency is a difficult task, a challenge that requires new strategies to unearth optimized compounds. We present a theory of thermoelectric transport in electron doped SrTiO$_{3}$, based on a realistic tight binding model that includes relevant scattering processes. We compare our calculations against a wide panel of experimental data, both bulk and thin films. We find a qualitative and quantitative agreement over both a wide range of temperatures and carrier concentrations, from light to heavily doped. Moreover, the results appear insensitive to the nature of the dopant La, B, Gd and Nb. Thus, the quantitative success found in the case of SrTiO$_{3}$, reveals an efficient procedure to explore new routes to improve the thermoelectric properties in oxides.  
\end{abstract}

\maketitle

In the context of critical energy and environmental issues, there has been a recent increase of interest in thermoelectric (TE) materials, which have the property to convert waste heat into electricity \cite{Snyder,Disalvo,Dresselhaus,Nolas,book,Heremans}.The efficiency of thermoelectric conversion depends on the value of the dimensionless figure of merit $ZT=\frac{S^{2}\sigma T}{\kappa}$ where $S$ is the Seebeck coefficient, $\sigma$ the electrical conductivity, T is the absolute temperature and $\kappa$ the thermal conductivity, usually dominated by phonon scattering processes. So far, a large majority of efforts to improve ZT have focused on reducing the lattice thermal conductivity by enhancing phonon scattering by processes such as alloying \cite{Steele,Cahill}, anharmonicity \cite{Garg}, or even by introducing nanoinclusions/inhomogeneties into the bulk matrix \cite{Faleev,Wang,Zhang}. Clearly, further optimization of TE properties will require an enhancement of the numerator of the figure of merit called the thermoelectric power factor PF=$S^{2}\sigma$, thus an increase of the Seebeck coefficient while maintaining a high electrical conductivity \cite{Tritt}. Improving the thermoelectric figure of merit ZT is one of the greatest challenges in material science. Among the wide family of interesting TE materials such as skutterudite, Heusler alloys, clathrates, binary tellurides or topological insulators, oxides could be a good alternative. Indeed, these compounds exibit other interesting properties such as being environment friendly, they contain abundant elements, they are cheap, resistant and stable up to high temperature. The perovskite material SrTiO$_{3}$ (STO) is particularly interesting because it has already a relatively large power factor of the order of 20 $\mu W/cm\cdot K^{2}$ comparable to that of the best known TE materials such as Bi$_2$Te$_3$ \cite{Okuda2001}. However, because of its relatively high thermal conductivity of $\kappa\approx$ 11 W/m$\cdot$K \cite{Muta}, the ZT of STO is only 0.1. It is thus clear that the nanostructuration of the material (reduction of the thermal conductivity) combined with a judicious dopant could further boost the PF and thus lead to large values of ZT. 
In this work, we propose a detailed theoretical study of the TE transport in STO and compare our results with a wide panel of experimental published data for bulk and thin films. To complete the latter, we present our thermoelectric measurements on heavily La doped STO films epitaxially grown by MBE on STO (001) substrate \cite{Vila,Mihai}.

First principles studies show that the lowest conduction bands ($\pi^{*}$) in STO have mainly the Ti d character \cite{Mattheiss,Wolfram,Shanti,Usui}. Therefore, instead of performing full ab initio calculations, our strategy consists in building up a minimum tight-binding (TB) Hamiltonian from the most relevant electronic bands. This allows more general discussions and facilitates the identification of the relevant underlying mechanisms just by tuning a single well-defined physical parameter. First, we define the minimal but realistic TB Hamiltonian for the t$_{2g}$ orbitals and then we introduce the relevant scattering processes needed to address the TE properties beyond the constant relaxation time approximation. 

The Hamiltonian reads, $\hat{H}=\hat{H_{0}}+\hat{H}_{dis}$ where,
\begin{eqnarray}
\hat{H_{0}}=\sum_{\footnotesize\textbf{ij},\alpha\beta} t^{\alpha\beta}_{\footnotesize\textbf{i}\textbf{j}} c_{\footnotesize\textbf{j}\beta}^{\dagger}c_{\footnotesize\textbf{i}\alpha}
\end{eqnarray}

\begin{eqnarray}
\hat{H}_{dis}=\sum_{\footnotesize\textbf{i},\alpha} \epsilon_{\footnotesize\textbf{i}} c_{\footnotesize\textbf{i}\alpha}^{\dagger}c_{\footnotesize\textbf{i}\alpha}
\end{eqnarray}
$\hat{H_{0}}$ is the TB part and $\hat{H}_{dis}$ describes the effects of disorder (dopants substitution and intrinsic defects).
$\left|\alpha \right>$, $\left|\beta \right>$ denote $\left|xy \right>$, $\left|yz \right>$ or $\left|zx \right>$, the 3 t$_{2g}$ d-orbitals of Ti. The integrals $t^{\alpha\beta}_{\footnotesize \textbf{i}\textbf{j}}$ are restricted to nearest and next nearest neighbour only. We also assume no hopping between d-bands, e.g. $t^{\alpha\beta}_{\footnotesize\textbf{ij}}=0$ if $\alpha\ne\beta$. Resulting from the symetry of the orbital, we have for the d$_{xy}$-band the following set of hoppings: in-plane nearest neighbour t$_{1}$, out of plane nearest neighbour t$_{2}$ and in-plane next nearest neighbour t$_{3}$ (the out of plane hopping is negligible). The parameters are t$_{1}$=0.277 eV, t$_{2}$=0.031 eV and t$_{3}$=0.076 eV as estimated in Ref. \onlinecite{3dtbmodel}. The other two bands (d$_{xy}$ and d$_{zx}$) are obtained by applying a circular permutation (x,y,z)$\rightarrow$(y,z,x)$\rightarrow$(z,x,y). The TB Hamiltonian becomes, $\hat{H_{0}}=\sum_{\textbf{k},\alpha} \epsilon^{0}_{\alpha}(\textbf{k}) c_{\footnotesize\textbf{k}\alpha}^{\dagger}c_{\footnotesize\textbf{k}\alpha}$ where $\epsilon^{0}_{xy}(\textbf{k})=-2t_{1}\left(cos(k_{x}a)+cos(k_{y}a)\right)-2t_{2}cos(k_{z}a)-4t_{3}cos(k_{x}a)cos(k_{y}a)$, where the lattice parameter $a=3.9$~$\angstrom$ in STO. The on-site scattering potentials $\epsilon_{\footnotesize\textbf{i}}$ in $\hat{H}_{dis}$ are chosen randomly within a box distribution of width W. The treatment of $\hat{H}_{dis}$ is discussed in what follows.

\begin{figure}
\includegraphics[scale=0.4,angle=0]{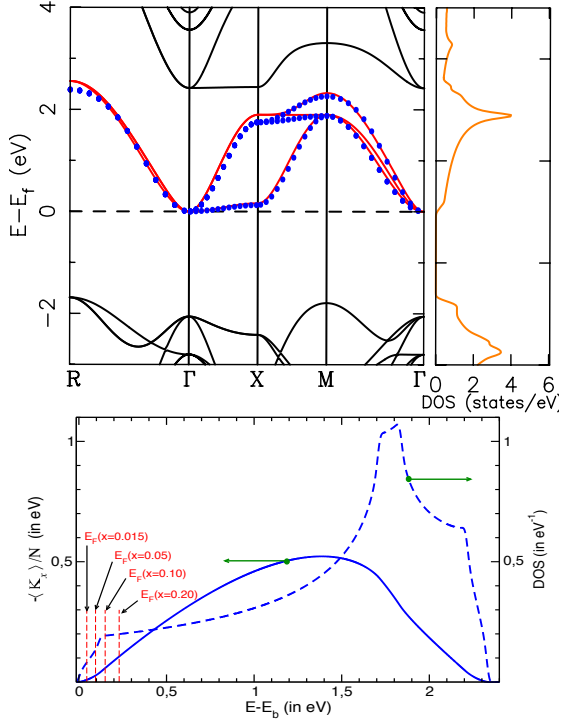}
\caption{\label{fig1} (Top) Density of states and dispersion obtained with both Siesta
and within the minimal tight binding model for the 3 t$_{2g}$ bands (blue dots).
(Bottom) Tight binding model calculations of (i) the normalized density of states (blue dashed line, right axis) and (ii) the reduced Drude weight $-\langle K_{x}\rangle/N$ (blue continuous line, left axis) as a function of E. The dashed vertical lines indicate the position of the Fermi level for dopant concentration ranging from x=0.015 to x=0.20.
}
\end{figure}

The conductivity and the Seebeck coefficient are, 
\begin{eqnarray}
\sigma(\mu,T)=-\int \Sigma(E,T)\frac{\partial f(E,\mu)}{\partial E}dE
\end{eqnarray}
\begin{eqnarray}
S(\mu,T)=\frac{1}{eT\sigma(\mu,T)}\int \Sigma(E,T)(E-\mu)\frac{\partial f(E,\mu)}{\partial E}dE
\end{eqnarray}
where $\mu$ is the T-dependent chemical potential and $\Sigma(E,T)=D(E)\tau(E,T)$, D(E) is the Drude weight calculated at T=0 K. By analogy with the classical Drude formalism, one can write $D(E)=\frac{ne^{2}}{m_{t}}$ where $n$ is the carrier density and $m_{t}$ the transport effective mass. $\tau(E,T)$ is the energy and temperature dependent quasiparticle lifetime.

D(E) is the order parameter for the metal-insulator phase transition, and can be directly extracted from the following sum rule \citep{Kohn,Millis,Scalapino,Bouzerar},
\begin{eqnarray}
D(E)= -\frac{2}{\pi}\int_{0}^{+\infty}\sigma_{reg}(\omega,E)d\omega -\frac{\sigma_{0}}{N\hbar}\langle \hat{K_x} \rangle(E)
\label{eqD} 
\end{eqnarray}
where $\sigma_{reg}$ is the regular (incoherent) part of the optical conductivity, $\sigma_{0}=\frac{e^2}{\hbar a}= 6258~\Omega^{-1}\cdot cm^{-1}$, N is the number of sites and $\hat{K_x}=-\frac{\partial^{2}\hat{H}}{\partial \kappa_{x}^{2}}$ ($\kappa_{x}=k_{x}a$).

In this study, we restrict ourselves to weak disorder regime, a justified approximation for samples exhibiting a good metallic behaviour. This regime corresponds to $k_{F}l_{e} \gg$ 1, where $k_{F}$ is the Fermi wave vector and $l_{e}$ the mean free path. As will be seen, this is indeed the case for most samples considered here. The localisation effects expected to play a crucial role at low temperature for low doping (typically below 1-2\% in La doped STO) are currently under investigation \cite{Simon}.
In the weak disorder regime, D(E) is reduced to the second term in equation (\ref{eqD}), since the transfer of weight from the Drude peak to finite frequencies is small, hence $D(E) \approx -\frac{\sigma_{0}}{N\hbar}\langle \hat{K_x} \rangle(E)$. Note also that $D(E)$ is dominated by d$_{xy}$ and d$_{xz}$ bands that contribute equally, whilst d$_{yz}$ band has a negligible contribution  (the hopping in the $x$-direction is very small).
\begin{figure}
\includegraphics[scale=0.26,angle=0]{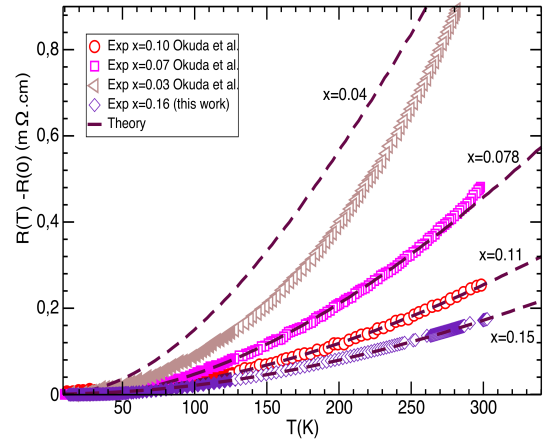}
\caption{\label{fig2} Resistivity as a function of temperature: theory (dashed lines) vs experiments (symbols). The data points on bulk and thin films are taken from 
ref.\onlinecite{Okuda2001} and our measurements. In the theoretical calculations, the concentration of dopant is directly indicated in the figure. 
}
\end{figure}
We now briefly discuss the nature of the scattering rate. It has two contributions: $\frac{1}{\tau(E,T)} = \frac{1}{\tau_{dis}(E)}+\frac{1}{\tau_{th}(T,E)}$.
$\tau_{dis}(E)$ denotes the effect of disorder resulting from the cationic substitutions and presence of other defects (intrinsic, dislocations, grain boundaries) whilst $\tau_{th}(T,E)$ is the temperature dependent part. Its origin is electron-phonon processes (e-ph) and electron-electron (e-e) scattering. In oxides such as STO, several studies showing a T$^2$ dependent resistivity suggest that the e-e mechanism dominates over the e-ph contribution up to relatively large temperatures \cite{Baratoff,VanderMarel1,Mikheev,Klimin}. Thus, we consider this term only. Using the Fermi golden rule we get $\frac{\hbar}{\tau_{dis}(E)}=2\pi \langle \epsilon_{\footnotesize\textbf{i}}^{2} \rangle \rho(E) = \frac{\pi W^2}{6} \rho(E)$ where $\rho(E)$ is the density of states. The thermal contribution has the form, $\frac{\hbar}{\tau_{th}(E)}=C\frac{(k_{B}T)^{2}}{E-E_{b}}$ where C is a dimensionless constant and E$_{b}$ the energy at the bottom of the conduction band. There is no simple and direct way to estimate C, it depends on the Thomas-Fermi screening length scale, carrier concentration and topology of the Fermi surface. Below, we explain the procedure that allows to set free parameters (C,W).

Fig.~\ref{fig1}(top) shows the dispersion obtained from both (i) ab-initio calculations (Siesta) and (ii) TB model described previously.
The Siesta calculations are performed using the d$\zeta$p basis and GGA \cite{siesta}. We find an excellent agreement between both calculations, that fully supports the 3 t$_{2g}$ bands Hamiltonian modelization.
In Fig.~\ref{fig1}(bottom) we have also plotted both the calculated DOS and the reduced Drude weight $-\langle K_{x}\rangle/N$ as a function of E. It can be seen that beyond 10 $\%$ doping D(E) increases almost linearly. The corresponding Fermi energy coincides with that of the kink in the DOS or edge of the heavy electron bands as plotted in Fig.~\ref{fig1}(top). Below 10 $\%$, we find $D(E) \propto (E-E_b)^\frac{5}{3}$, in contrast with the free electron model for which the power is 3/2.

\begin{figure}
\includegraphics[scale=0.26,angle=0]{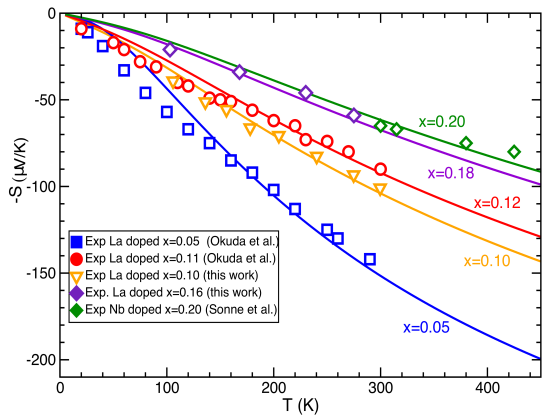}
\caption{\label{fig3}Seebeck coefficient S as a function of temperature: theory vs experiment. The squares and circles are data extracted from ref. \onlinecite{Okuda2001} and the green diamonds from ref. \onlinecite{Sonne}. Both triangles and purple diamonds have been obtained in the present study. The continuous lines are the calculated values. The dopant concentration is indicated in the figure.}
\end{figure}

In Fig.~\ref{fig2} the resistivity is plotted as a function of temperature. We set the free parameters C and W in such a way that we reproduce the resistivity measurements for $10 \%$ La doped sample of Ref.\onlinecite{Okuda2001} corresponding to a carrier density of $n=x/a^3=1.7 \,10^{21} \;cm^{-3}$. More precisely W is set to reproduce R(T$=0$ K) and C adjusted to give R(T$=300$ K) measured experimentally. The motivation for this choice is the good metallic behaviour found for this concentration of dopant. This leads respectively to $W=0.17$ eV and $C=$24.5. These parameters are now set for the whole study. This value of W is consistent with the assumption of weak disorder regime, indeed $W \ll W_b$ where the bandwidth of the conduction band $W_b$ is of the order of 2 eV. We now discuss the results of our calculations. First, for sufficiently large doping ($ x\ge 6-7\%$) we observe a very good quantitative agreement between theory and experiment for the whole range of temperature. Below (between 3-5$\%$) some deviations at low temperature between theory and experiment are visible. Theory leads to slighly larger values of the resistivity, the reasons for this could be manifold. First, the simplicity of the model: the low energy band structure (below the kink in the DOS) should be improved. Secondly, the electron-electron scattering rate used here does not include the true nature of the d-orbitals. In other words, the non spherical nature of the Fermi surface resulting from the strong hopping anisotropy is not taken into account. Finally, the presence of native defects, such as oxygen vacancies, not included here, should also have an effect. At much lower concentration below 1.5 $\%$, it is experimentally observed that the resistivity increases strongly \cite{Okuda2001}. A possible explanation could be that at low density the Fermi level gets closer to the mobility edge (separating extended from localized states). Localization effects, not included in the present study, should lead to a strong suppression of the Drude weight (significant transfer of weight to the regular part of the conductivity) and thus to an increase of the resistivity. If we further decrease the carrier density, we expect a metal to insulator transition below a critical concentration as seen in Ref.\onlinecite{Okuda2001}.

In Fig.~\ref{fig3} we have plotted the measured Seebeck coefficient (S) as a function of temperature in both La and Nb doped samples, together with the theoretical calculations.
For large concentration (beyond 5$\%$), we observe an overall good quantitative agreement between theory and experiments.
At lower concentration, the agreement is very good above 100 K, and below this temperature the experimental data slightly deviates from the calculations.
This larger measured |S| could be a consequence of the Fermi level proximity to localized states region. This feature is expected to become more pronounced as the carrier density is further reduced, leading eventually to a minimum in |S| at low temperature. This is for instance observed in 1.5$\%$ La doped samples in ref.\onlinecite{Okuda2001} and \onlinecite{Stemmer}. A well defined minimum is clearly seen in both papers when the electron density is small enough. Such a minimum is often attributed to phonon-drag. However, the relevance of this mechanism is still highly debated. Thus, it would be of great interest to clarify for low doped STO, whether the minimum is a signature of Anderson localization or due to phonon drag.

\begin{figure}
\includegraphics[scale=0.28,angle=0]{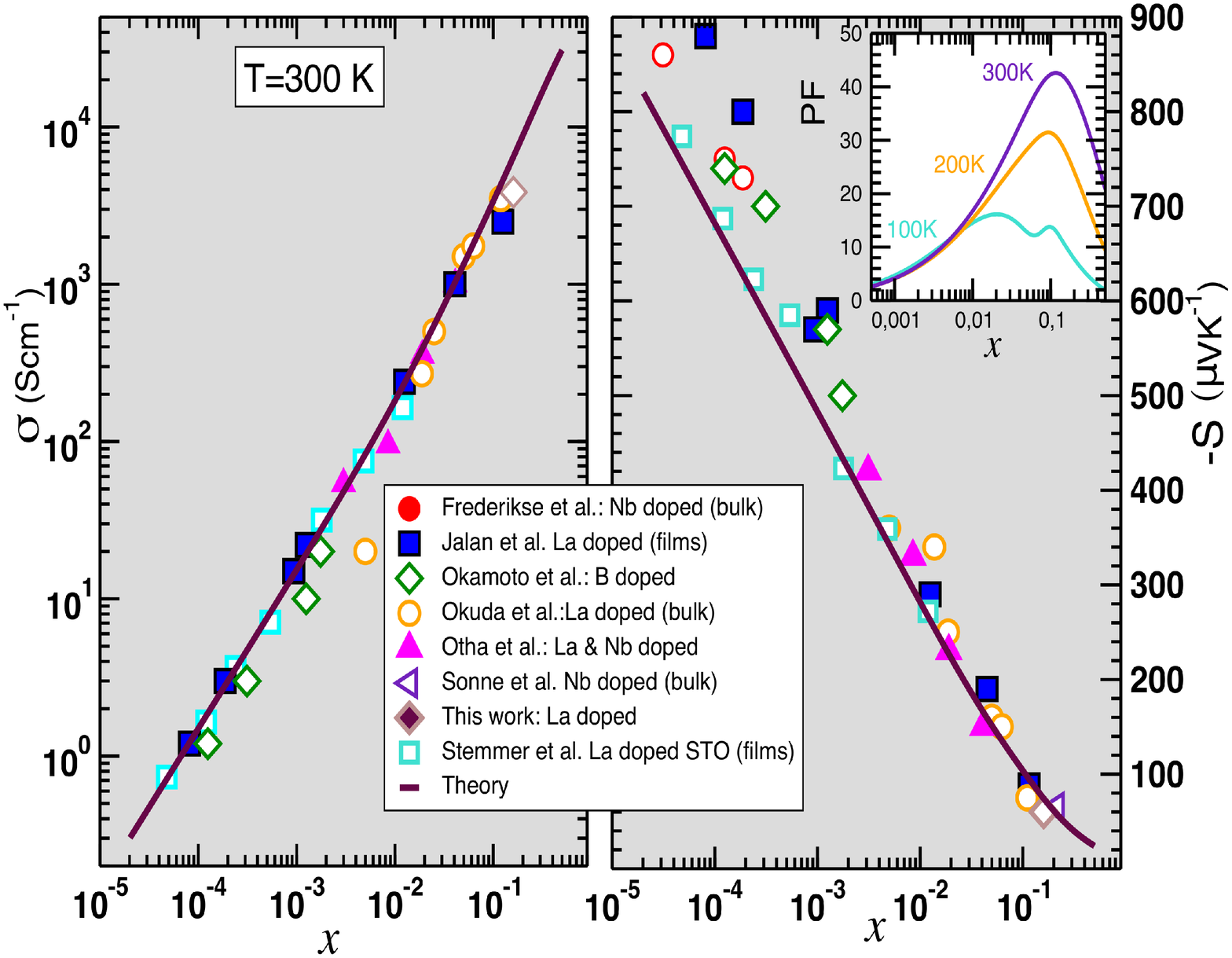}
\caption{\label{fig4}
Conductivity and Seebeck as a function of the doping ($x$) at $T=
300$ K. From very dilute up to 20$\%$ doped: Theory vs experiment. The
inset shows the  calculated power factor PF (in $\mu W cm^{-1}K^{-2}$) for 3 different temperatures. Experimental data (symbols) have been extracted from
ref. \onlinecite{Okuda2001,Sonne,Okamoto,Jalan,Frederikse,Otha,Stemmer} and from our measurements. The nature of the dopant and material (bulk or film) is indicated in the figure.
}
\end{figure}
We propose now to compare the calculated carrier dependent electric conductivity and Seebeck coefficient at $T=300$ K to available experimental data. The results are depicted in Fig.~\ref{fig4}.
The agreement found between theory and experiments is very good for carrier density spreading over four decades (extremely low to heavily doped).
The agreement is almost insensitive to the electron donor (La, Nb $\&$ B) and to the nature of the sample (bulk or thin films). The conductivity varies over four decades and is impressively well reproduced by the theory. This is especially surprising considering the simplicity of our realistic TB model.
\begin{figure}
\includegraphics[scale=0.27,angle=0]{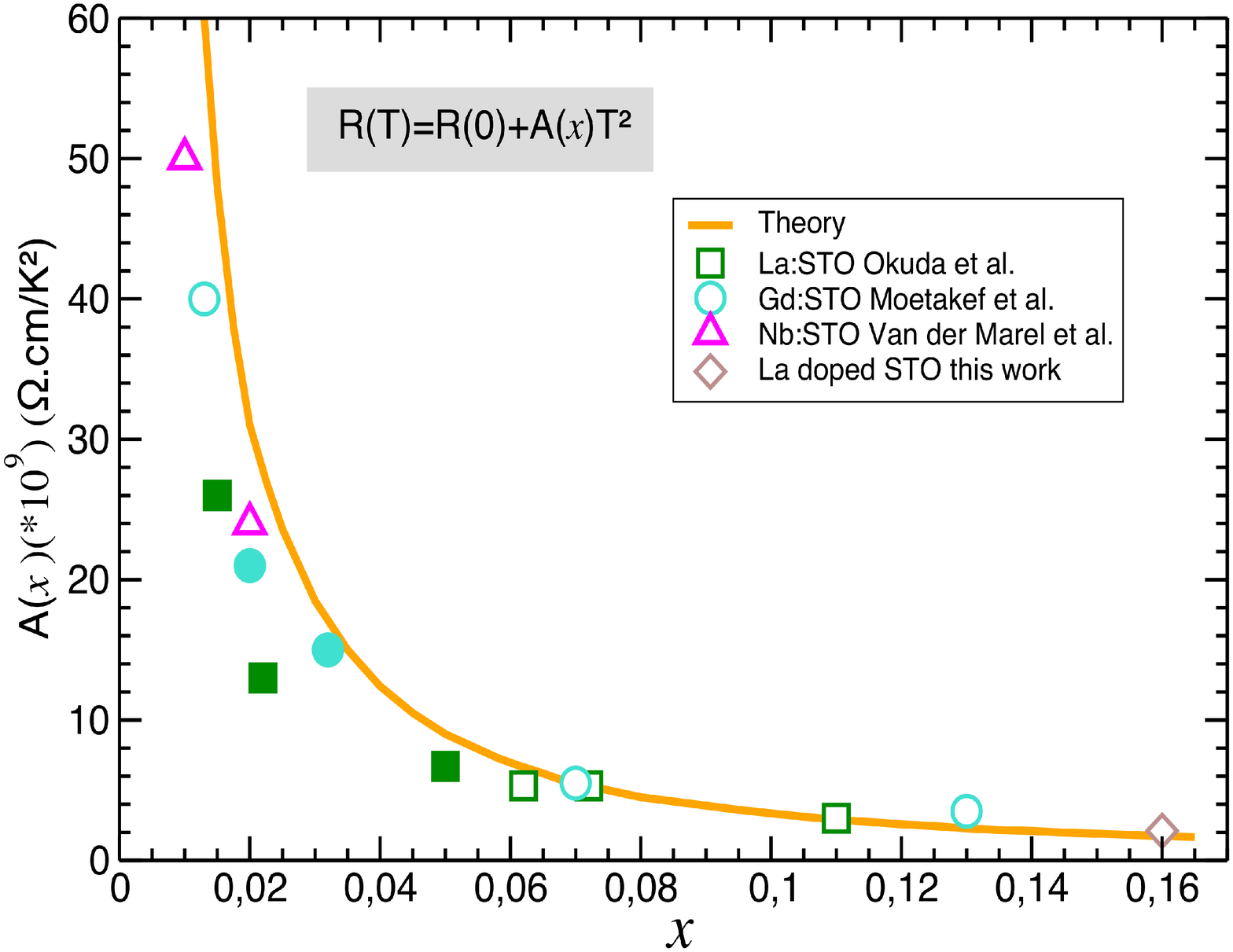}
\caption{\label{fig5} Resistivity coefficient A as a function of the electron density $x$: Theory and experiment. Experimental data have been extracted from ref. \onlinecite{Okuda2001,Moetakef,VanderMarel} for La, Nb, and Gd doped STO and our measurements on La doped STO films .
}
\end{figure}
The Seebeck coefficient varies from -60 $\mu$V$\cdot$K$^{-1}$ at about $x$=0.20 to -900 $\mu$V$\cdot$K$^{-1}$ for $x$=10$^{-5}$. The quantitative agreement is again very good for the overall range of carrier concentrations and weakly depends on the dopant and nature of the sample. A deviation can be seen below $x$=0.001, the measured S are more dispersed but slightly higher than those calculated by about 10-15\%. This small deviation could be attributed to localization effects. It is important to remember that as the temperature is reduced the effect of localization should become more pronounced.
In the inset, we have plotted the calculated power factor (PF) as a function of $x$ for three different temperatures.
First, we observe a maximum located at $x\approx 0.1$ (for T=300 K) that progressively shifts towards lower concentrations as the temperature is decreased.
The PF increases significantly with the temperature. This results from a stronger increase of S$^{2}$ that overcompensates the reduction of the conductivity.
At T= 300 K, we obtain a relatively high value for the power factor PF=43 $\mu W /cm\cdot K^{2}$ that is in good agreement with recent measurements in La doped thin films
\cite{Jalan}. A secondary peak in PF is observed for the lowest temperature, this is attributed to the kink in the DOS (see Fig.~\ref{fig1}).
This figure illustrates nicely the universal character of the present theoretical approach for the TE properties in STO. It is important to stress that, at T=300 K and over the whole range of carrier densities, our calculations reveal that the scattering rate is controlled by C only. Thus, the conductivity is inversely proportional to C and the Seebeck coefficient is independent on both parameters. Thus, the crucial ingredients are (i) the realistic band structure and (ii) the T$^2$/E dependence of the e-e scattering rate.

In the last section we discuss the T dependence of the resistivity (in metallic samples only). Due to the e-e scattering mechanism, the resistivity can be accurately fitted by $R(T)=R(0)+AT^2$. Indeed, it has been shown recently by Lin et al. \cite{Linetal} that the T$^2$ law persists down to very low concentration of dopants. Similar experimental results have been reported as well in Ref.\onlinecite{Mikheev2}. In Fig.~\ref{fig5} we plot the variation of A with the electron density, but our concern here is the comparison between theory and experiment.
Thus, it is sufficient to restrict ourselves to data ranging from intermediate to heavy doping. In Fig.~\ref{fig5}, we find that the agreement is very good above $x=0.03$. Beyond 8$\%$ doping, the variation of A with respect to $x$ is very weak (almost flat). Below 3-4$\%$, A varies very strongly as $x$ is reduced, and  a deviation from the measured values is observed. Note however, that the data become very dispersed as seen for instance in the 2\% doped compounds. It should also be mentioned that the measured carrier concentrations are not precisely known which could also contribute to the observed deviation. In addition, at low carrier densities, the details and nature of the disorder may play a role. From this figure we can conclude that the overall agreement is rather good.

To conclude, in this study that combines theory and experiments we have addressed the thermoelectric properties in electron doped SrTiO$_{3}$.
Our theory based on a realistic 3 bands tight binding model that includes relevant scattering processes (weak disorder and e-e scattering mechanism) captures qualitatively and quantitatively well the electronic transport properties in these compounds. The agreement found between theory and experiments covers a wide range of concentrations, from very low to heavily doped.
The results are weakly sensitive to the dopant La, Nb, B and even Gd and to the nature of the material, thin films or bulk.
The calculations show that STO can already exhibit a relatively high power factor of 43 $\mu W\cdot /cm\cdot K^{2}$ at room temperature for about 10$\%$ doping. This is in good agreement with recent experimental data. This study provides an efficient procedure to explore new pathways to improve the thermoelectric properties in oxides and other families of compounds. It should also facilitate the search for new dopants and allow for including effects such as nanostructuration and localization.

SP, RB and MA acknowledge the EU for funding through the project TIPS (H2020-ICT-02-2014-644453) and the French National Agency for funding the project MASCOTH (ANR-13-PRGE-0004). RB and MA acknowledge P. Regreny and G. St Girons for MBE support.

\end{document}